%% file: NGC-4696-12-09_coronal_submitted_2.tex
\documentclass[useAMS,usegraphicx,usenatbib]{mn2e} 
\usepackage{amsmath,graphicx}
\usepackage{ulem}
\usepackage{color}
\usepackage{graphicx}
\usepackage{times} 
\usepackage{amssymb}
\usepackage{amsmath}
\usepackage{lscape}
\usepackage{url}
\usepackage{multirow}
\input{defn}
\newif\ifAMStwofonts
\AMStwofontstrue
\begin{document}

\title[Coronal line emission in NGC 4696] {Detection of optical coronal emission from 10$^{6}$~K gas in the core of the Centaurus cluster} \author[R.E.A.~Canning et al.] {\parbox[]{6.in}
  { R.~E.~A.~Canning$^{1}$\thanks{E-mail:
      bcanning@ast.cam.ac.uk}, A.~C.~Fabian$^{1}$, R.~M.~Johnstone$^{1}$, J.~S.~Sanders$^{1}$, C.~S.~Crawford$^{1}$, N.~A.~Hatch$^2$ and G.~J.~Ferland$^3$\\ } \\
  \footnotesize
  $^{1}$Institute of Astronomy, Madingley Road, Cambridge, CB3 0HA\\
  $^{2}$University of Nottingham, School of Physics \& Astronomy, Nottingham NG7 2RD\\
  $^{3}$Department of Physics, University of Kentucky, Lexington KY 40506, USA}

\maketitle

\begin{abstract} 
We report a detection ($3.5\times10^{37}\pm5.6\times10^{36}$~\ergps) of the optical 
coronal emission line [Fe {\small X}]$\lambda$6374
and upper limits of four other coronal lines using
high resolution VIMOS spectra centred on NGC 4696, the brightest cluster
galaxy in the Centaurus cluster. Emission from these
lines is indicative of gas at temperatures between 1 and 5 million K so traces the
interstellar gas in NGC 4696. The rate of cooling derived from the upper limits is consistent
with the cooling rate from X-ray observations ($\sim$10~\Msunpyr) however we detect twice 
the luminosity
expected for [Fe {\small X}]$\lambda$6374 emission, at 1 million K, our lowest
temperature probe. We suggest this emission is due to the gas being heated rather
than cooling out of the intracluster medium. We detect no coronal lines from [Ca {\small XV}],
which are expected from the 5 million K gas seen near the centre in X-rays with
\chandra. Calcium is however
likely to be depleted from the gas phase onto dust grains in the central regions of NGC 4696.
\end{abstract}


\section{Introduction}
The X-ray emitting gas at the centers of clusters of galaxies can have 
very short cooling times. Without a heating source this gas should cool and 
condense at rates of up to thousands of solar masses per year. However, high resolution
X-ray spectroscopy has been unable to detect large 
quantities of gas cooling below one third of the cluster virial 
temperature (for a review see \citealt{peterson2006}). Observed rates
of star formation and quantities of cool and cold gas in brightest cluster
galaxies (BCGs), the most massive galaxies known, are far lower than expected 
(see for example \citealt{mcnamara2006}). For the hot gas to be radiating 
but not cooling in the predicted quantities a heating mechanism is required.

To determine how the heating and cooling balance in clusters of
galaxies is maintained requires a multi-wavelength approach to probe the many
phases of matter. The hot intra-cluster medium (ICM), seen in X-rays, 
is at a temperature of $\sim10^{7}-10^{8}$~K. Cool, optical 
emission line gas, found to surround the majority of BCGs in cool core 
clusters is seen at temperatures of $\sim10^{4}$~K. Cold molecular gas
reservoirs have also been observed, however observations of gas at 
intermediate temperatures ($10^{5}-10^{6}$~K)
have thus far proved elusive.

High-ionisation, collisionally-excited optical coronal lines are emitted 
at temperatures between $10^{5}-10^{6}$~K making them important tracers 
of intermediate temperature gas. Optical emission lines 
have the advantage of being observable from the ground. The high spectral
and spatial resolution in principle allow us to determine and trace the 
velocity structure, and therefore examine the transport processes, in
the gas. Emission lines from these plasmas can also provide a direct means of 
measuring the mass flow rate \citep{cowie1981} and therefore the rate at which 
the ICM is condensing. \cite{graney1990}, \cite{sarazin1991} and \cite{voit1994} have 
modelled coronal line emission in conditions appropriate to those of the cores
of galaxy clusters.

There have been several attempts to detect coronal line emission
from cooling gas in clusters of galaxies, however reports of a significant
detection remain inconclusive.
\cite{hu1985} observed the core regions of 14 clusters
of galaxies but were unable to detect the presence of the [Fe {\small XIV}]$\lambda$5303
in any of their long slit spectra. Similarly \cite{heckman1989} attempted
detection of another coronal line, [Fe {\small X}]$\lambda$6374, in a sample of nine
suspected `cooling flow' clusters. Likewise they found no detectable
emission.

\begin{figure*}
 \begin{center}
  \includegraphics[width=0.4\textwidth]{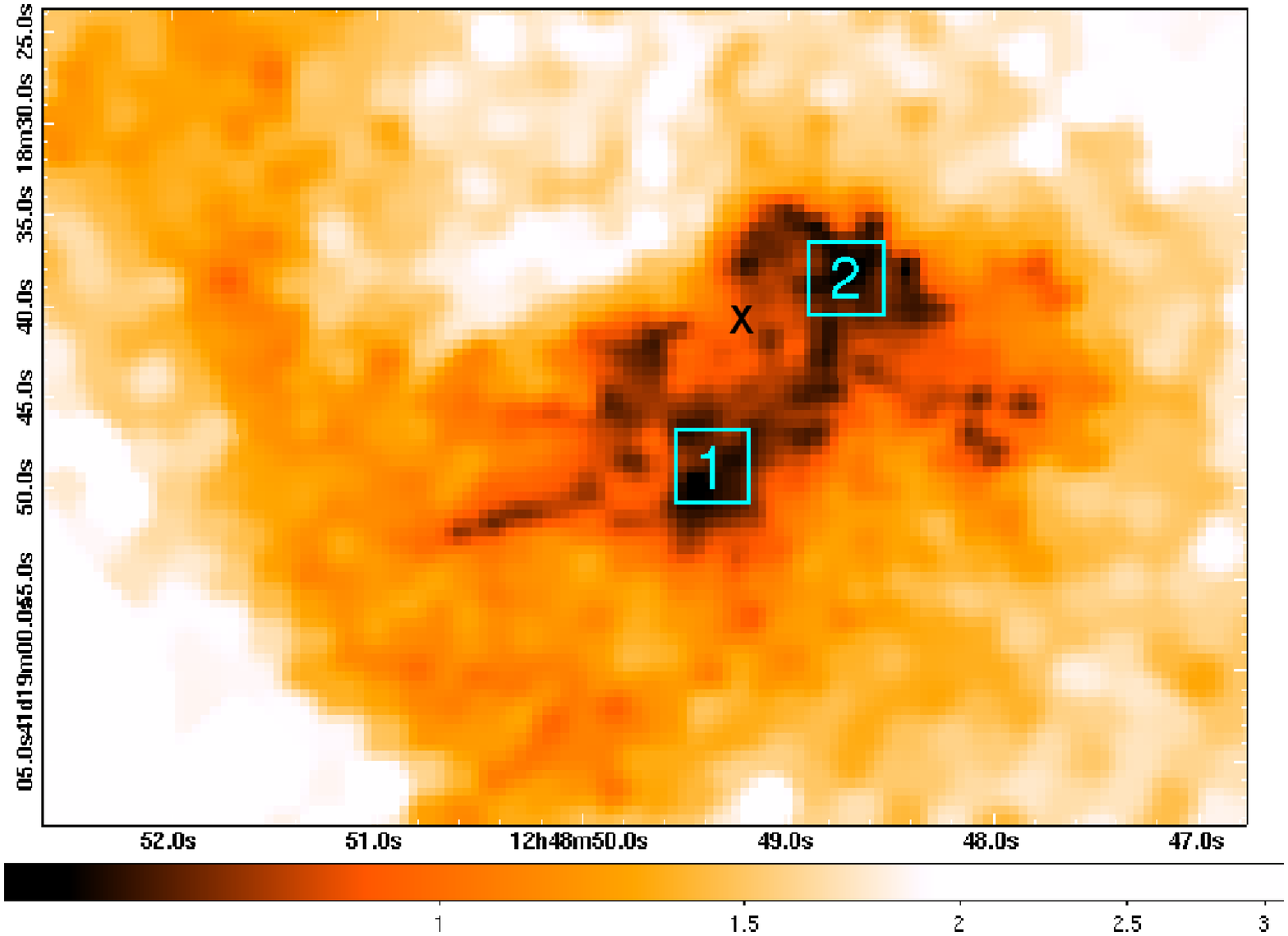}
  \includegraphics[width=0.4\textwidth]{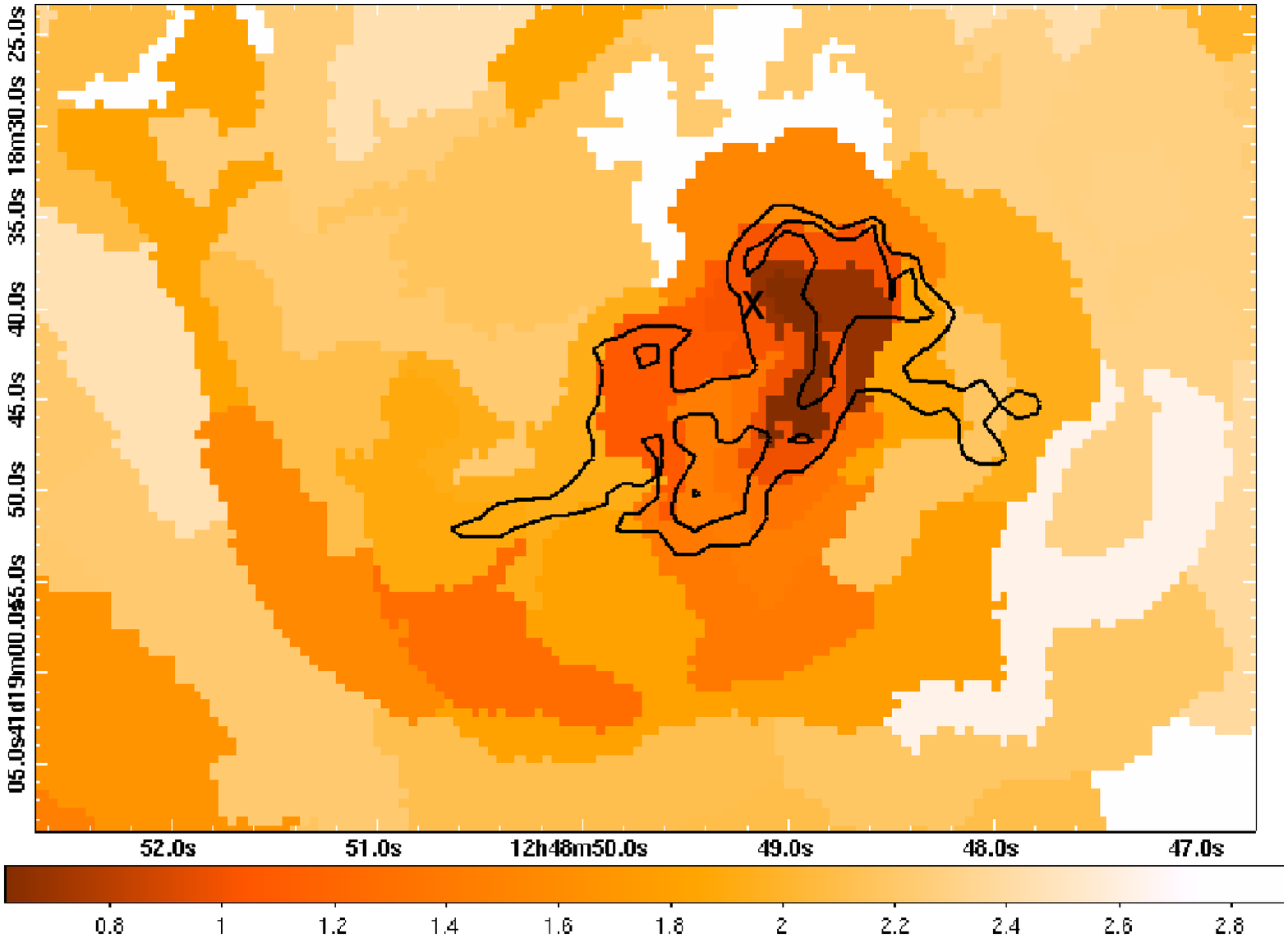}
  \includegraphics[width=0.4\textwidth]{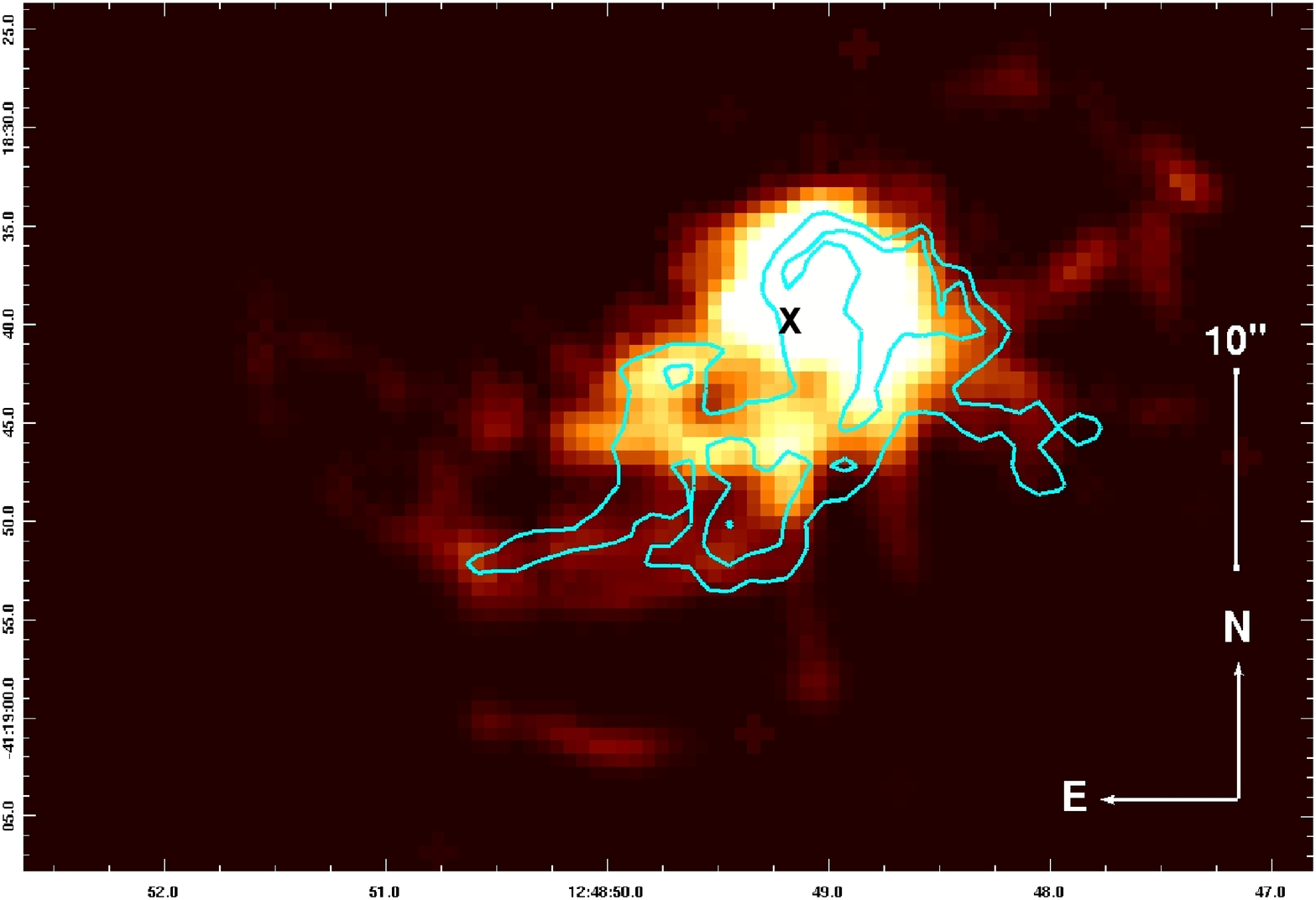}
\caption{Top left: X-ray temperature map of the centre of the Centaurus cluster \protect \cite{fabian2005b}. 
The colour bar is in units of \keV. Box 1 and 2 are situated at the regions of 
lowest X-ray temperature. Top right: X-ray metallicity map based on Fe abundance. 
The colour bar is in units of solar metalicity. The $0.4-0.8$~\keV\ X-ray temperature contours 
are overlaid. Bottom: [N {\small II}]$\lambda6583$ emission in NGC 4696 overlaid with X-ray 
contours of $0.4-0.8$~\keV\ gas. All images are on the same scale
(10'' $=2.1$\kpc). The black X indicates the centre of the radio emission. Box
1 is at position RA $12^{h}48^{m}49^{s}.4$, Dec. $-41^{\circ}$ 18' 48.9'' (J2000) and
box 2 is situated at RA $12^{h}48^{m}48^{s}.7$, Dec. $-41^{\circ}$ 18' 38.6'' (J2000). Both boxes are
5 by 5 fibres large corresponding to 3.4 arcseconds (0.7~\kpc) across. \label{maps}}
 \end{center}
\end{figure*}

The first reported detection of coronal emission was by \cite{anton1991} in Abell 1795.
The authors claimed to have found a 13$\sigma$ detection of the [Fe {\small X}]$\lambda$6374
line in the core of the cluster with an aperture centred on the central
2'' of the cD galaxy.
A later study of the same object by \cite{donahue1994} failed to confirm this 
detection. The sample
used by \cite{donahue1994} contained five massive cool core clusters and in this
same paper they report three more non-detections (A2199, 2A0335+096 and A2597) and one 
marginal (3$\sigma$) detection in PKS 0745-191. This detection has also 
failed to be confirmed in a later study by \cite{yan1995}.
   
The emission line nebulosities around NGC 1275, the BCG
in the Perseus cluster were studied by \cite{shields1992}. They searched for
[Fe {\small X}]$\lambda$6374 emission in off-nuclear regions, away from the central AGN, but also
obtained no detection.

Up until now the two most recent searches for coronal line emission have been by
\cite{yan1995} and \cite{sabra2000}. \cite{yan1995} searched for emission from
[Fe {\small X}]$\lambda$6374 in Abell 1795 and PKS 0745-191, the only two clusters where 
detections of coronal line emission had previously been reported. Contrary
to these results they did not find any significant detection in either cluster but 
were able to place upper limits on the surface brightness a factor of 10 deeper.

In a larger program studying the properties of the emission line nebulosity in
NGC 1275, \cite{sabra2000} searched for coronal lines over a wide
wavelength range using Keck. Six coronal line transitions were covered in this range from
[Ni XII]$\lambda$4232 to [Ni XII]$\lambda$6702. No coronal line emission was seen
in their data.

This work reports on deep, high spectral resolution, integral field spectroscopy (IFS)
observations of NGC 4696 (RA $12^{h}48^{m}49^{s}$, Dec. $-41^{\circ}$ 18' 40'' J2000), the BCG in the Centaurus cluster, Abell 3526. The 
Centaurus cluster at redshift $z=0.0104$ is the second nearest 
example of a cool core cluster. The heating and cooling in this cluster
is apparently very well balanced despite the short central cooling time of only
200~\Myr. NGC 4696 houses a radio source, and multiple bubbles.
These are accompanied 
by soft X-ray filaments and a sharp rise in the metal abundance in the central 
30~\kpc, among the highest seen in any cluster, $\sim$~twice solar 
\citep{sanders2002, fabian2005b, sanders2006}. Centaurus also has the broadest range 
of X-ray temperatures seen, containing gas from 0.35 to 3.7~\keV, over a 
factor of 10 in temperature \citep{sanders2008b}.

\cite{crawford2005} presented images showing the extensive, 
filamentary H$\alpha$ nebulosity surrounding NGC 4696. These 
share the morphology of the soft X-ray filaments, and of a prominent dust lane
\citep{sparks1989}. The low X-ray temperatures, short cooling times and 
exceptionally high metallicities in this nearby cluster make it the ideal target 
for a deep search for coronal line emission.

The observations are briefly described in $\S$ 2, method, analysis and 
limits on coronal line emission is given in $\S$ 3 and in $\S$ 4 we discuss 
the implications of these limits. We summarise our results and main 
conclusions in $\S$ 5. 
At the redshift of the Centaurus cluster ($z=0.0104$, 44.3~\Mpc) one \asec\ corresponds 
to 0.210~\kpc\ (throughout this paper we adopt H$_0=71$~\kmpspMpc, $\Omega_{\mathrm{M}}=0.27$ and 
$\Omega_{\Lambda}=0.73$). Abundances given in this paper are relative to the 
solar metallicities of \cite{anders1989}.

\section{Observations and Data Reduction}

Observations were made on 2009 March 27-30 using the VIsible MultiObject Spectrograph 
(VIMOS) on the VLT in La Paranal, Chile (see \citealt{lefevre2003} and \citealt{zanichelli2005} for a 
description of the VIMOS IFU and a discussion of data reduction techniques). We obtained High 
Resolution Orange (HRO) data using VIMOS Integral Field Unit (IFU). We used the larger 0.67'' fibres 
giving a field of view of 27''$\times$27'' with the HR grism.

We acquired 10.5 hours exposure centred on the inner region of NGC 4696 (12$^{\mathrm{h}}$48'49.3'', -41$^{\circ}$18'40''),
the deepest observations thus far to try to detect emission from cooling hot gas in these objects. We also
took six 15 minute exposures using the same set up to image the entire galaxy including many
sky fibres.

\begin{figure}
 \begin{center}
  \includegraphics[width=0.5\textwidth]{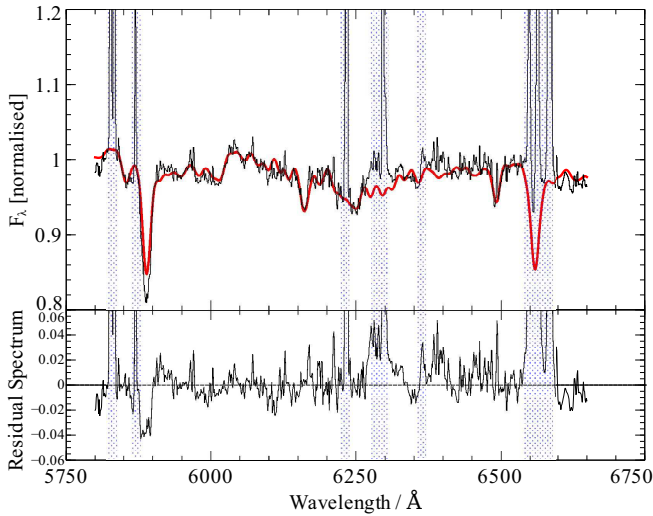}
  \caption{The best-fit BC03 SSP model (red) to the observed spectrum, from
box 1, black between $5700-6700$~$\mathrm{\text{\AA}}$ and the
fit residuals. Regions where we expect object emission lines and regions of poor sky subtraction
were masked out in the fit and are shown above in grey. \label{ssp}}
 \end{center}
\end{figure}

\begin{figure}
 \begin{center}
  \includegraphics[width=0.5\textwidth]{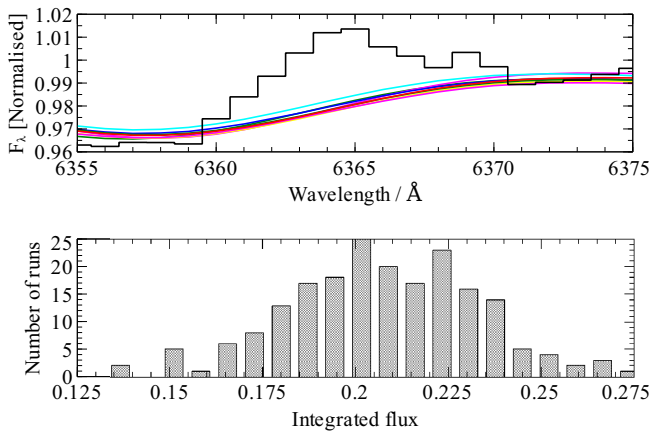}
  \caption{Top: The range of \Starlight\ fits over the [Fe {\small X}]$\lambda$6374 region,
in box 1, determined by perturbing and re-fitting the observed spectrum.
Bottom: The distribution of integrated flux values for 200 \Starlight\ fits over the 
[Fe {\small X}]$\lambda$6374 region. \label{ssperr}}
 \end{center}
\end{figure}

The data were reduced by the VIPGI\footnote{VIPGI-VIMOS Interactive Pipeline Graphical Interface, 
obtained from http://cosmos.iasf-milano.inaf.it/pandora/.} pipeline (see \citealt{scodeggio2005} for a 
description of VIPGI). The 3D-datacubes were combined and analysed with a set of IDL routines (R. 
Sharp, private communication).

Instrument flexures in VIMOS are very dependent on rotator position and suffer 
from hysteresis \citep{Amico2008}. For this reason we took calibration frames after 
each observation block.

Cosmic-ray rejection, final fibre-to-fibre transmission corrections, sky subtraction,
correction for Galactic extinction and shifting to the object rest frame
were performed outside of VIPGI using IDL routines. The data reduction procedure will
be explained thoroughly in a forthcoming paper, Canning et al. (in prep), detailing 
observations of the emission line nebulae surrounding NGC 4696.

\section{Method, Analysis, Detections and Limits}

Our HR Orange spectra cover the wavelengths of a number of coronal lines, specifically 
[Fe {\small XIV}]$\lambda5303$, [Ca {\small XV}]$\lambda5445$, [Ca {\small XV}]$\lambda5694$, [Fe {\small X}]$\lambda6374$
and [Ni {\small XV}]$\lambda6702$.

The expected location of the [Fe {\small X}]$\lambda$6374 feature coincides with
the red wing of the [O {\small I}]$\lambda$6363 object emission so requires special
attention (see section \ref{sec:FeX}). None of our other expected coronal lines coincide 
with object emission lines or sky lines.

\subsection{Continuum subtraction}
\label{sec:subtraction}

\begin{figure*}
 \begin{center}
  \includegraphics[width=0.8\textwidth]{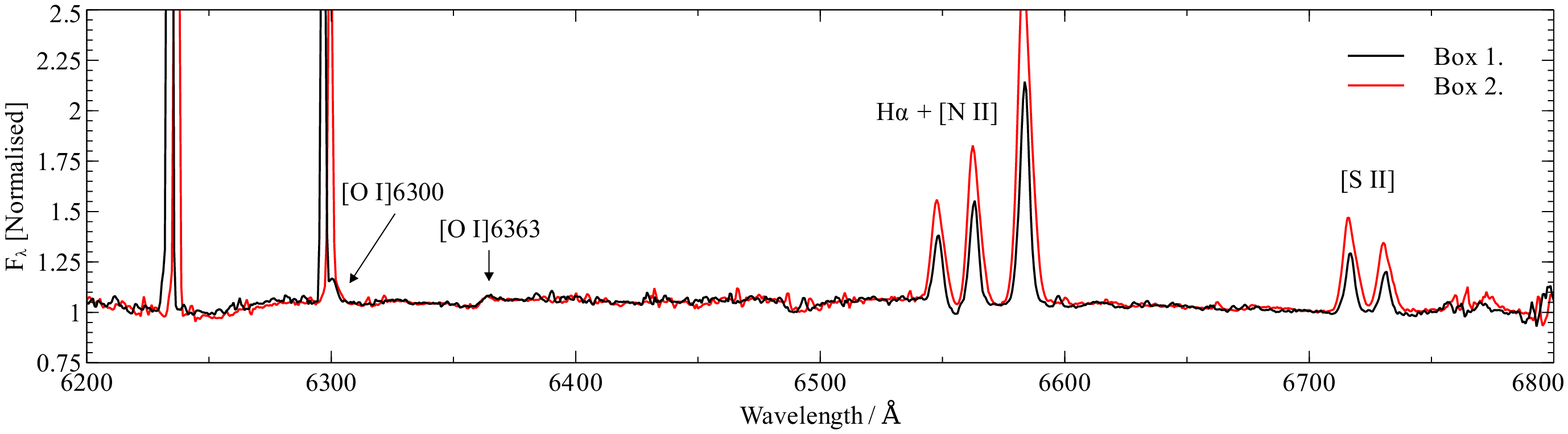}
  \includegraphics[width=0.4\textwidth]{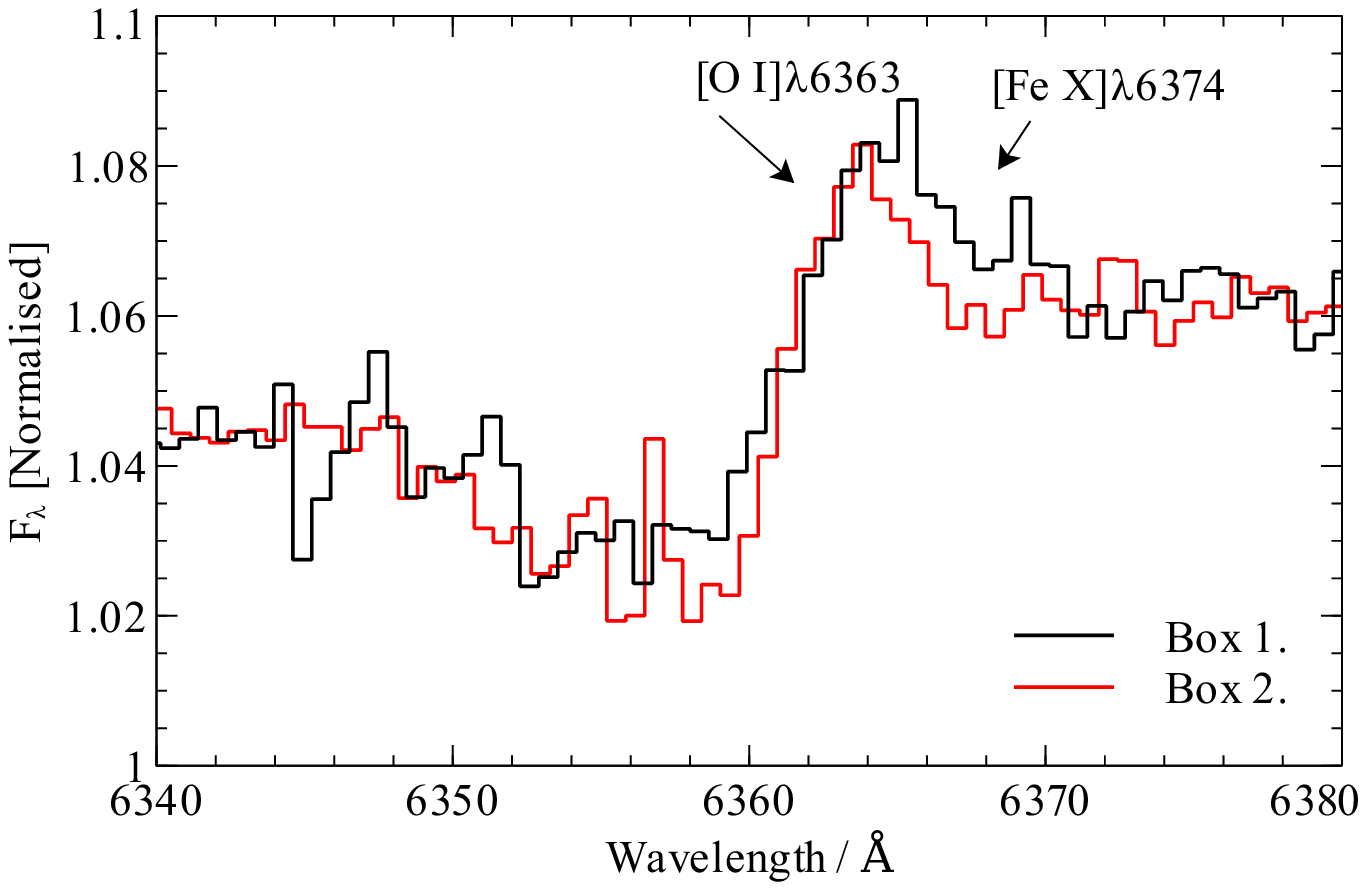}
  \caption{Top: Spectra from box 1 (black) and box 2 (red) without continuum subtraction. All
sky lines except [O {\small I}]$\lambda$6300 and [O {\small I}]$\lambda$6363 sky emission have
been subtracted in order to more clearly see object emission line features. Bottom: A blow-up of the expected [Fe {\small X}]$\lambda$6374 region.
Emission from cool gas in box 2 has a broader velocity width than box 1 in general except over the 
[O {\small I}]$\lambda$6363 and [Fe {\small X}]$\lambda$6374 region. This demonstrates the excess flux on the red wing
of oxygen emission in box 1. This can not be accounted for by subtraction of a template from the object
[O {\small I}]$\lambda$6300 emission line.\label{box1and2}}
 \end{center}
\end{figure*}

\begin{figure}
 \begin{center}
  \includegraphics[width=0.5\textwidth]{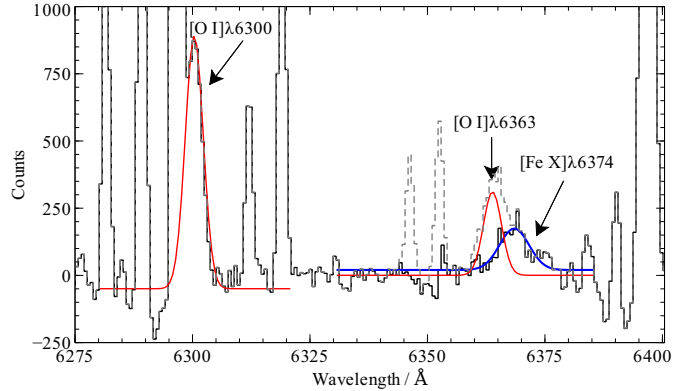}
  \caption{Spectrum from box 1, after subtraction of a smooth continuum, showing the fit to the [O {\small I}]$\lambda$6300 object emission
line and the scaled template for the [O {\small I}]$\lambda$6363 emission (red). The blue
Gaussian is the result of a fit to the remaining excess flux after subtraction of the 
oxygen line. The dashed spectrum is the original continuum subtracted data and the solid 
line shows the effect of the [O {\small I}]$\lambda$6363 line subtraction. Here also two neighbouring
sky lines are fit and subtracted. The continuum is allowed to vary from zero and is fit locally
around the emission lines of interest. \label{fit}}
 \end{center}
\end{figure}

When determining flux from possibly broad and faint emission lines, errors in 
continuum subtraction are likely to dominate. For this reason continuum 
subtraction was attempted in a number of ways. The first method we employ is to
use a small sample of 
fibers from within the galaxy, in a region where no significant coronal 
emission is seen. A second method is to subtract a smoothed continuum from the spectra 
(as in \citealt{donahue1994}) and finally we model the continuum with simple
stellar populations (SSP) models from \cite{bc03}, hearafter refered to as BC03.
These methods are described below.

1. In an initial search we binned the spectra based on X-ray temperature, the 
lowest temperature regions (box 1 and 2, Fig. \ref{maps}) overlap with the 
temperature range at which some coronal lines are emitted, specifically 
the lines of [Ca {\small XV}] (5 million K).
In box 1 we found evidence for a flux excess on the red wing of the [O {\small I}]$\lambda$6363
emission line, blueshifted slightly from the 6374~$\mathrm{\text{\AA}}$ where we would expect
[Fe {\small X}]$\lambda$6374. We found
no significant excess in box 2. We then use the spectrum 
of box 2 to correct the continuum of box 1 and search for significant 
flux excess in regions of expected coronal emission.

2. A smooth continuum is fitted to the sky subtracted spectra with 
all regions where emission lines are expected and regions of poor sky
subtraction masked out. The smooth continuum is then subtracted from those spectra.

\begin{figure}
 \begin{center}
  \includegraphics[width=0.5\textwidth]{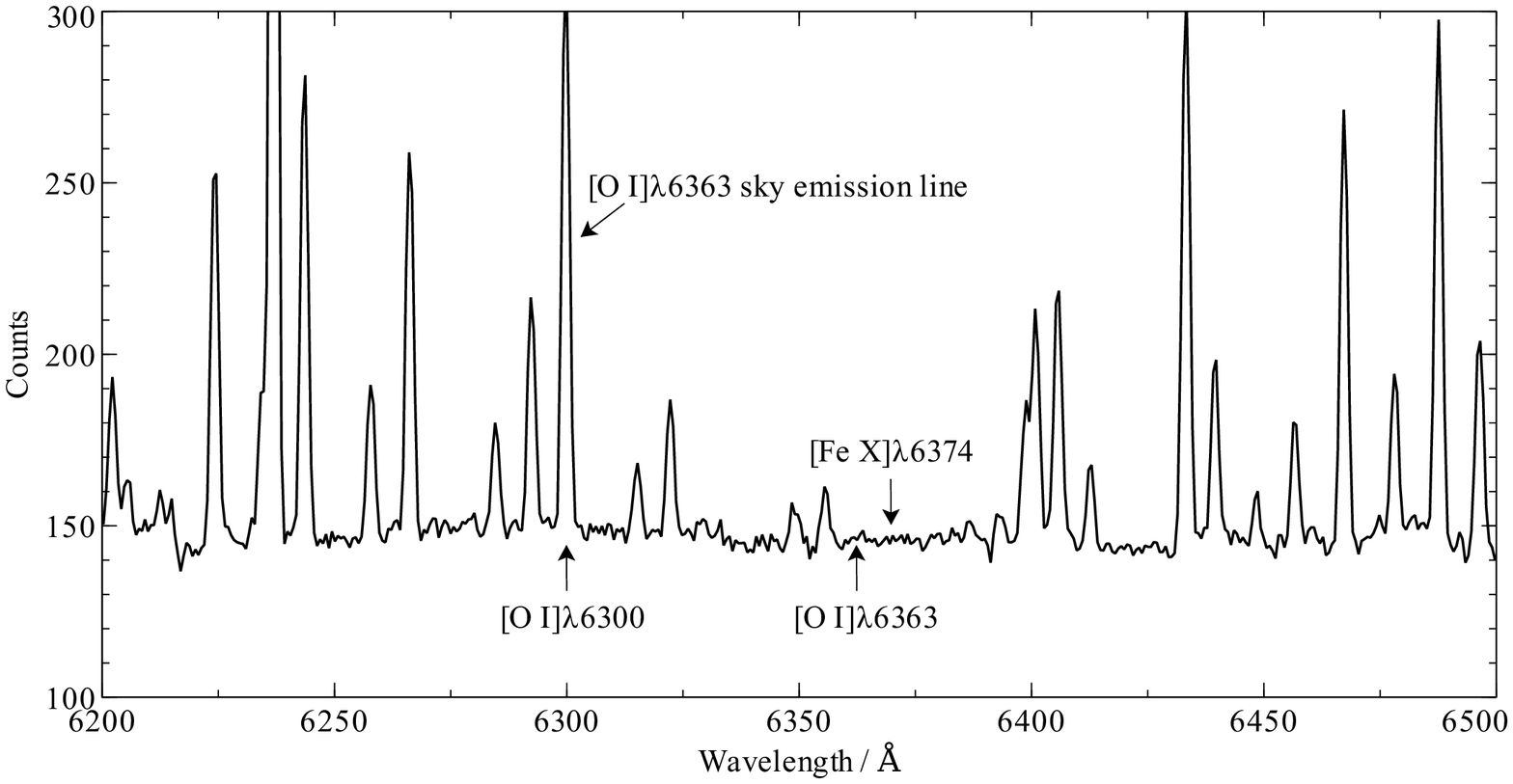}
\caption{The sky spectra extracted from fibres at the edge of the field $\sim$~35'' from the
centre of NGC 4696. The [O {\small I}]$\lambda$6363 sky emission line is labelled as are the
regions where we would expect the object [O {\small I}] emission lines and the [Fe {\small X}]
emission. The spectrum has been shifted to the rest-wavelength of NGC 4696. \label{sky}}
 \end{center}
\end{figure}

\begin{figure}
 \begin{center}
  \includegraphics[width=0.48\textwidth]{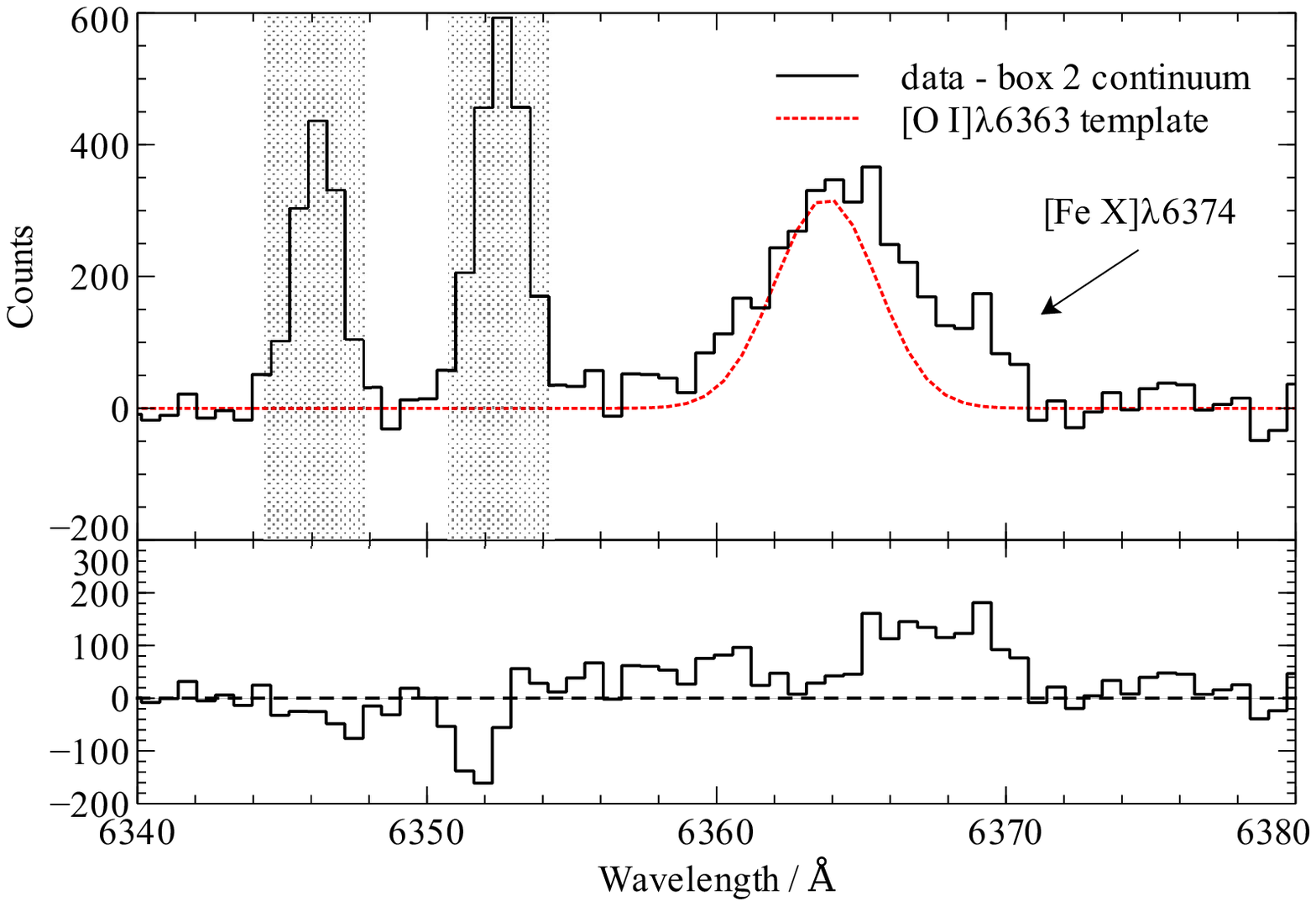}
  \includegraphics[width=0.48\textwidth]{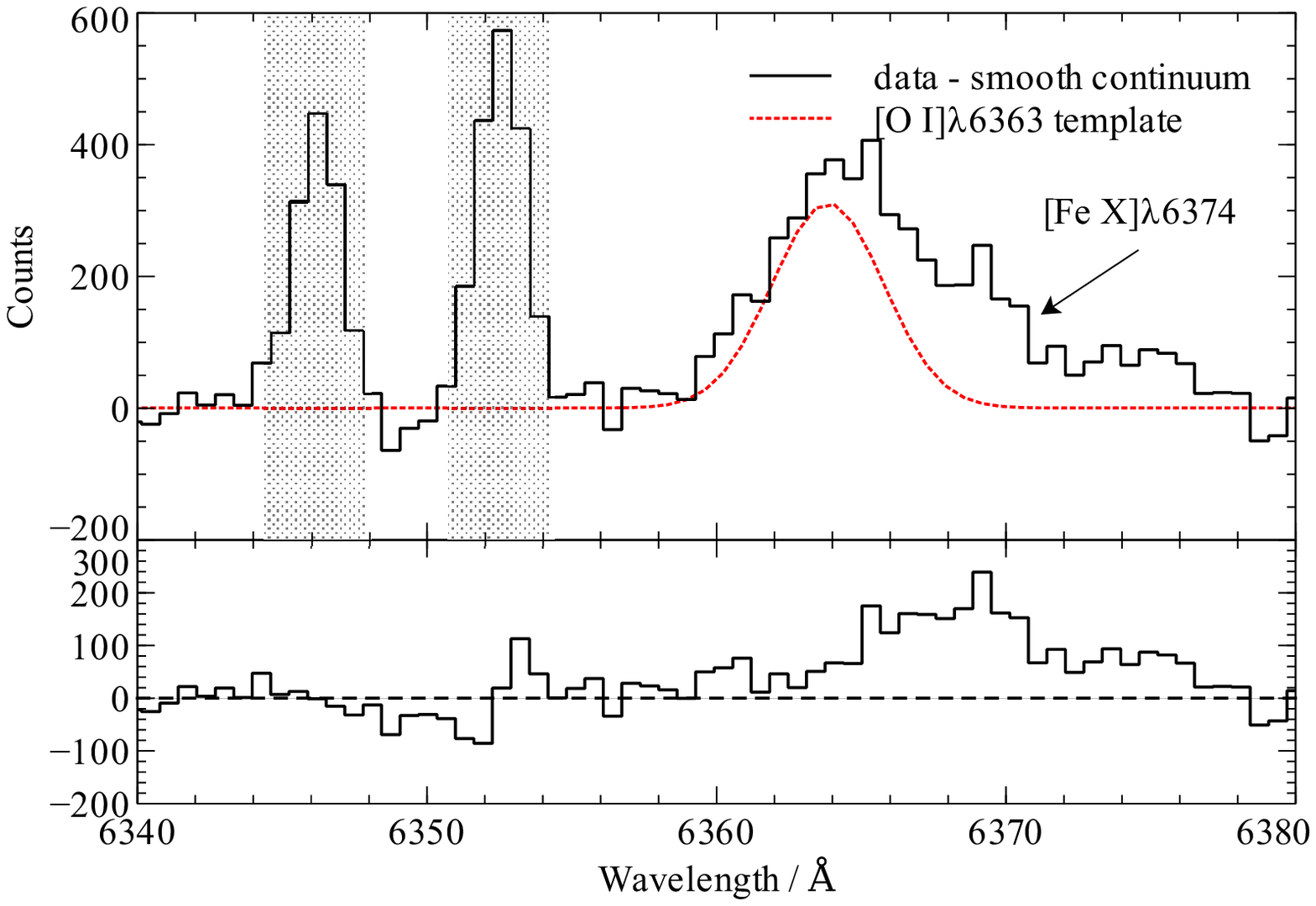}
  \includegraphics[width=0.48\textwidth]{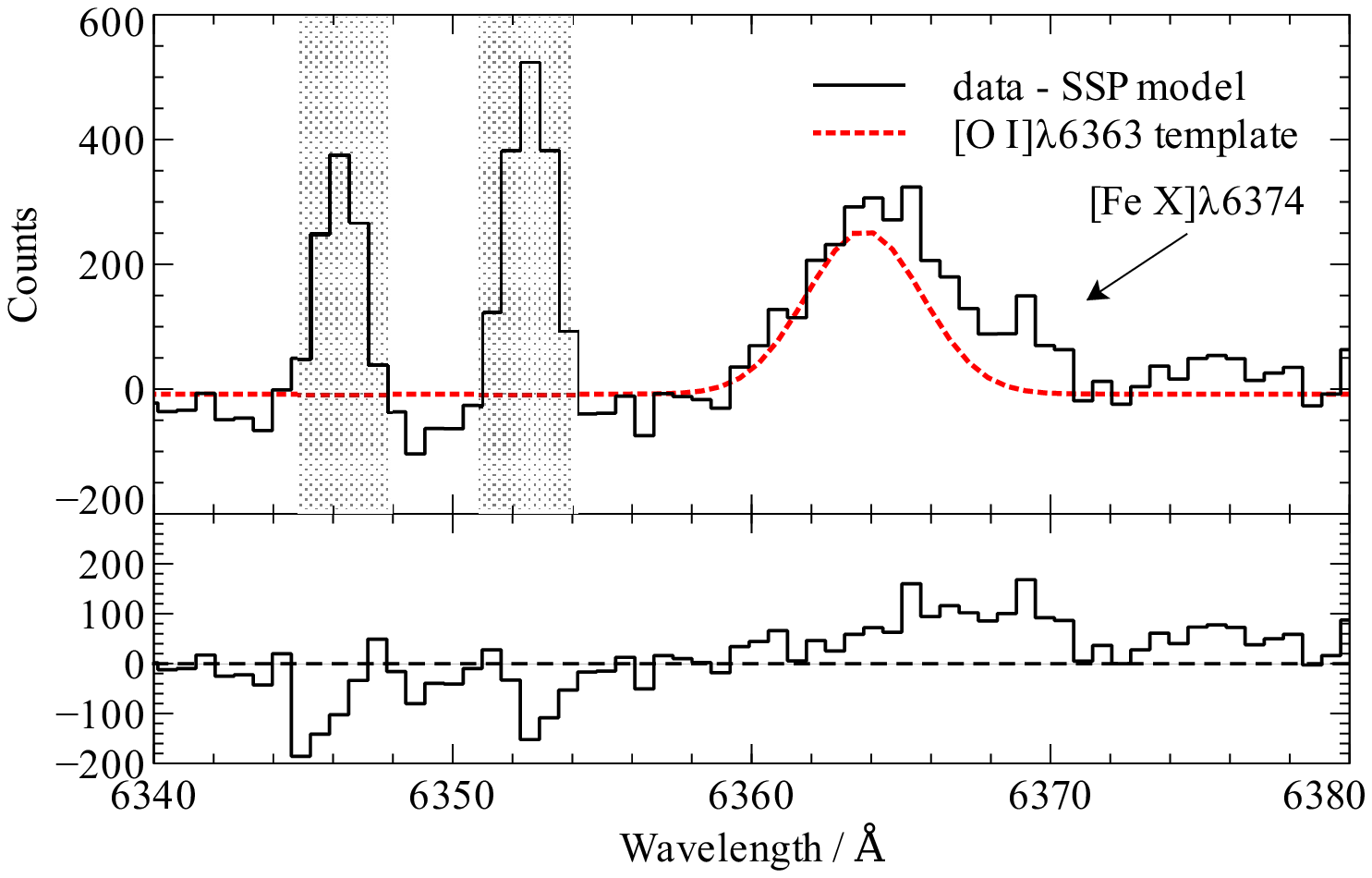}
\caption{Top: Box 1 spectra after continuum subtraction
using the continuum from box 2 spectra. 
Middle: Box 1 spectra with smooth continuum subtracted. 
Bottom: Box 1 spectra with
best fit \Starlight\ continuum model subtracted.
In each plot the top panel shows the data over 
the wavelength region containing the [O {\small I}]$\lambda$6363 object emission and probable 
[Fe {\small X}]$\lambda$6374 object emission. The red line shows the scaled [O {\small I}]$\lambda$6300 
emission used as a template to subtract the contribution of the [O {\small I}]$\lambda$6363 line.
The bottom panel, in each case, shows the residuals from a subtraction of the [O {\small I}]$\lambda$6363 
template. In each plot an excess of flux can be seen around a wavelength of 6368$\mathrm{\text{\AA}}$.
The two emission lines that have been greyed out, to the left of the [O {\small I}]$\lambda$6363 
line, are sky lines. These were fit and subtracted from the spectrum before a fit to the 
remaining excess flux was performed.\label{FeXfits}}
 \end{center}
\end{figure}

\begin{figure}
 \begin{center}
  \includegraphics[width=0.48\textwidth]{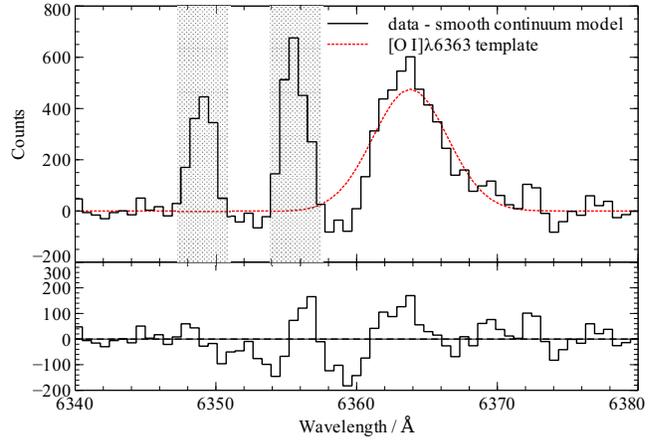}
\caption{Box 2 spectrum with a smooth continuum subtracted. There is no 
significant excess emission seen on the red wing of the object [O {\small I}]$\lambda$6363 
emission line. \label{TB2_fit}}
 \end{center}
\end{figure}

\begin{figure}
 \begin{center}
  \includegraphics[width=0.5\textwidth]{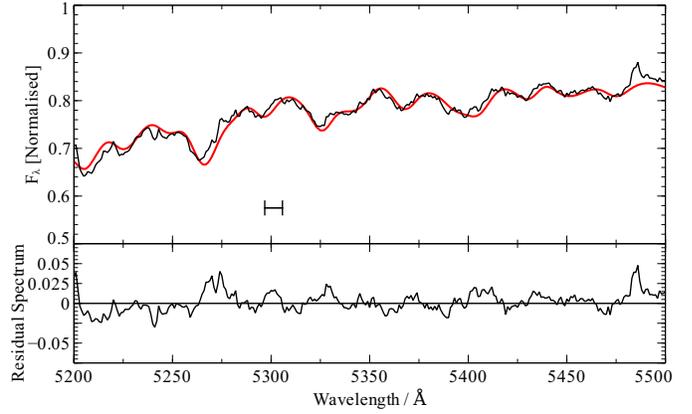}
\caption{The \Starlight\ fit in the bluest region of our spectra. After
subtraction of the model stellar continuum we are left with variations in 
the residual spectrum on the scale of $10-20~\mathrm{\text{\AA}}$. This is comparable
to the width of feature we find on the red wing of the [O {\small I}]$\lambda$6363
object emission line. For comparison the line shows the $\pm$1$\sigma$ width 
of a 475~\kmps\ FWHM velocity width line at the rest wavelength of the 
[Fe {\small XIV}]$\lambda$5303 coronal emission line. \label{blue}}
 \end{center}
\end{figure}

3. The continuum shape is fit using the \Starlight\ software \citep{cidfernandes2005, cidfernandes2009} 
and basefiles made up of 240 BC03 models spanning 6 metallicities, covering the
range $0.005-2.5~Z_{\odot}$, and 40 ages, covering the range $0-20$~Gyr. Using
this software requires that we re-grid our spectra to integer wavelengths.
\Starlight\ uses a Markov Chain approach to fitting the data. We determine the
parameters to use by fitting the data many times and looking at the distribution
of fits over the region of interest. The fit with the 
best $\chi^{2}$ value is taken as the continuum
and subtracted from the spectrum.
An example of a fit with emission lines masked out is shown in Fig. \ref{ssp}.
\Starlight\ does not allow us to examine the uncertainty in the parameters
from a fit and so we use a Monte Carlo approach to determine the error.
We quantify the additional uncertainty introduced due to the continuum subtraction 
by perturbing our data points by a Gaussian random number with 
mean and variance given by the data and associated poission error. These
new perturbed spectra are then
fit with the \Starlight\ software using the same parameters as before. We repeat this
200 times for our box 1 spectrum, the range of fits over the [Fe {\small X}]$\lambda$6374 region can
be seen in the top panel of Fig. \ref{ssperr}. The lower panel of Fig. \ref{ssperr}
shows the distribution of the integrated flux (counts-model) over the 
wavelength region $6355-6380~\mathrm{\text{\AA}}$. The Gaussian shape of this
distribution implies we have repeated the fits enough times to properly
sample the range of models which could provide a good fit to the data.
The 200 model continua are then re-gridded to the same wavelength
grid as our original data.
We estimate the additional uncertainty in each spaxel as the one sigma 
deviation in the range of models. This uncertainty is dependent on the error
spectrum of the entire wavelength range of the spectra being fit and as 
such will only be weakly correlated with the possionian error on each spaxel.
We therefore estimate the total 
uncertainty as the quadrature addition of the poisonion error and the 
uncertainty in the continuum model.

An advantage of using synthetic spectra for the continuum subtraction is that we
can ensure that there is no coronal line emission in the fit, so we are not subtracting
any signal. The stellar spectra are much more complicated than a smoothed profile
and we would like to test whether any apparent excess flux found where we would
expect coronal line emission could be due to a `bump' in the stellar spectra. However the 
spectral models assume the relative abundances are the same which may not be appropriate
to NGC 4696. The stellar continuum will also vary across the field and in binning spectra
over a large region of the galaxy we would expect the SSP model fit to deteriorate. For 
these regions fits were run seperately on all spectra i.e. we did not use the best fit 
\Starlight\ model of box 1 to correct for the stellar continuum of any spectra other
than box 1.

\subsection{[Fe {\small X}]$\lambda$6374}

\label{sec:FeX}

\begin{figure*}
 \begin{center}
  \includegraphics[width=0.8\textwidth]{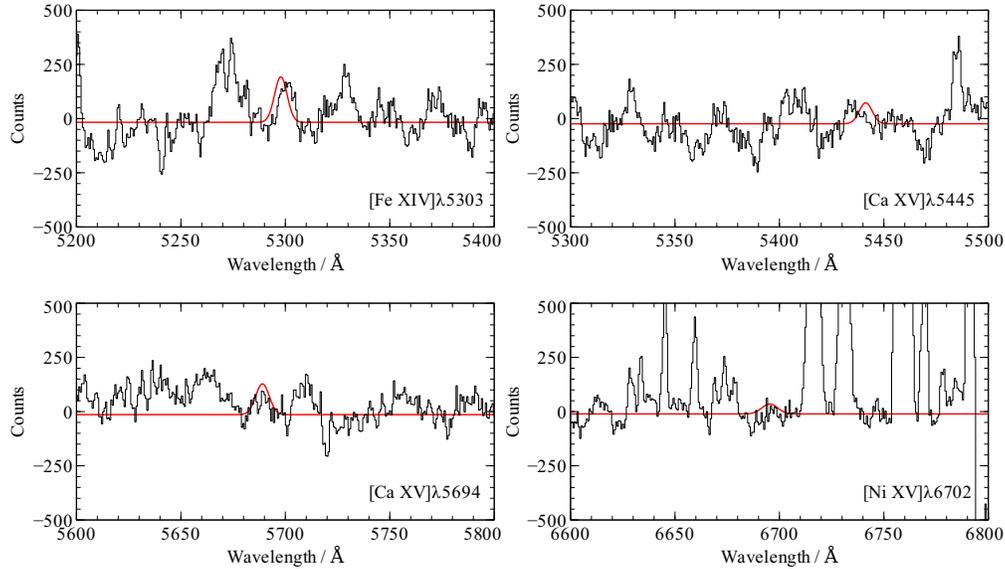}
\caption{The 90 per cent upper limits overplotted on the 
continuum subtracted spectra from box 1 for the lines of [Fe {\small XIV}]$\lambda5303$, 
[Ca {\small XV}]$\lambda5445$, [Ca {\small XV}]$\lambda5694$ and 
[Ni {\small XV}]$\lambda6702$. Here the continum was subtracted using SSP models
fit with the \Starlight\ software package. The redshift and velocity width of the
emission lines are constrained to be the same as that of the detected 
[Fe {\small X}]$\lambda$6374 line. \label{all}}
 \end{center}
\end{figure*}

The [Fe {\small X}]$\lambda$6374 coronal emission feature coincides with the red wing of
the [O {\small I}]$\lambda$6363 emission and as such warrants special treatment (see Fig. \ref{fit}). Unfortunately
the object [O {\small I}]$\lambda$6300 emission also coincides with the sky [O {\small I}]$\lambda$6363
emission feature. 

We simultaneously fit Gaussians to the lines of [O {\small I}], the [N {\small II}] doublet, 
the [S {\small II}] doublet, H$\alpha$ and the [Fe {\small X}] emission lines. This
is in order to allow for the errors in scaling and removing the [O {\small I}]$\lambda$6363
object emission line. A single Gaussian for each emission line in the obtical nebulosity 
gives a good fit in the regions of box 1 and 2 (see upper panel of Fig. \ref{box1and2}).

We tie the redshift and velocity width of the object optical emision 
line nebulae ([N {\small II}] doublet, [S {\small II}] doublet and 
H$\alpha$ and [O {\small I}]) emission lines. The redshift and velocity
width of the sky [O {\small I}] lines are also tied. The integrated
flux of the [N {\small II}] doublet and that of the two [O {\small I}]
lines can be tied to each other, the scaling in each case being dictated 
by atomic parameters \citep{osterbrock2006}. The integrated flux
of the [Fe {\small X}]$\lambda$6374 emission line is fixed to be a 
positive value but the redshift, and velocity width are allowed to be
free in the fit as there is no evidence yet for a spatial correspondence 
between the optical nebulosity and any coronal line emitting plasma. 
The continuum level for each line is fit in a region
local to the emission line. An example of the fitting process with
the fit to the object [O {\small I}] emission lines and the 
[Fe {\small X}]$\lambda$6374 emission line is shown in Fig. \ref{fit}.

\subsection{Detections and upper limits}

After continuum subtraction, detections and upper limits are established by fitting a Gaussian
line profile to the spectra using MPFIT \citep{more1978, markwardt2009}. The 
90 per cent flux limits are found by increasing the integrated flux in the fit until 
$\Delta\chi^{2}$ has grown by 2.7, these results are presented in columns 2-4
of table \ref{results}. The last three columns show, for two cooling rates
and metallicities, the predicted luminosity of
the coronal lines assuming they are the product of cooling from the hot ICM
(see \citealt{sarazin1991}).
During the fit we fix the Gaussian normalisation to be positive. When determining
the upper limits we also constrain
the redshift and velocity width to be the same as 
that of the [Fe {\small X}]$\lambda$6374 emission line. The coronal lines probe a 
broad range of temperatures so it is perhaps not obvious that they should have an
identical velocity. The continuum is fit
in a region local to the emission lines. 

We report a 6.3$\sigma$ detection of excess flux at the expected wavelength of
[Fe {\small X}]$\lambda$6374 emission in a region 15'' away from the 
nucleus (box 1, Fig. \ref{maps}). The quoted significance of this feature is from
a fit to the SSP model continuum subtracted spectrum. This method of continuum subtraction
gave the least significant result. We do not see
any evidence of this emission in a region of the galaxy where there is similar low
temperature X-ray gas nearer to the core (box 2). Fig. \ref{box1and2} shows a comparison of the spectra
in these two regions. The velocity width of the cool ($10^{4}$~K) gas near the core is 
larger than that of the gas farther away in all cases except the region around the
[O {\small I}]$\lambda$6363 object emission. Here the box 1 spectra shows an excess of flux
towards the red wing of the oxygen line, this is not seen in the box 2 spectra or in
the line profile of the [O {\small I}]$\lambda$6300 line in box 1.

Fig. \ref{sky} shows the sky spectrum. This spectrum is formed from regions at the edge of
the field of view where we have no detectable object emission. The regions where we would 
expect object [O {\small I}]$\lambda$6300, [O {\small I}]$\lambda$6363 and 
[Fe {\small X}]$\lambda$6374 emission is indicated. There is no obvious sky emission in
the region around the [O {\small I}]$\lambda$6363 and [Fe {\small X}]$\lambda$6374
lines and as the excess flux is seen as a very broad feature and it is not
seen everywhere in the spectra it is unlikely to be confusion due to sky emission.

To fit the excess flux
requires a redshift different from that of the optical emission line nebulae, the line
in box 1 is blueshifted with respect to the optical nebulosity. The line of
sight velocity difference between the redshift of the H$\alpha$ nebulae and the best fits
to the excess flux is $\sim$~230\kmps. This is comparable to the spread of the
line of sight velocities in the optical nebulosity itself.
The results from the three methods of continuum subtraction agree within 
the error (see Fig. \ref{FeXfits}). In box 1 the integrated flux with subtraction of 
the continuum using a region where no significant excess flux is 
observed, after subtraction of the [O {\small I}]$\lambda$6363 line, gives an integrated 
flux of $1.8\times10^{-16} \pm 2\times10^{-17}$\ergpcmsqps, subtraction of
a smoothed continuum gives an integrated flux in the line of 
$1.6\times10^{-16} \pm 2\times10^{-17}$\ergpcmsqps, in both these cases the error is 
determined from the possionian error in the spectrum which is similar
to the one sigma deviation after continuum subtraction of a nearby emission free region
of the spectrum, and finally 
SSP model fitting gives
$1.5\times10^{-16} \pm 2.4\times10^{-17}$\ergpcmsqps, including both the
possionian error and the additional error in the continuum fit. 
Fig. \ref{TB2_fit}
shows, for comparison, the same region with the template [O {\small I}]$\lambda$6363
line subtracted. There is no significant excess emission seen to the red-wing of
the [O {\small I}]$\lambda$6363 line.

In regions containing strong stellar continuum features where the
background spectrum is highly structured ($\lambda<6000~\mathrm{\text{\AA}}$)
the error
after continuum subtraction is larger than the simple poissonian error.

The noise features in these regions have 
a width similar to the emission feature we are looking for ($10-20~\mathrm{\text{\AA}}$, see Fig. \ref{blue}). We estimate
the noise in these regions in two ways. First we calculate the one sigma 
deviation from zero in an emission free region of the spectrum (after continuum subtraction)
on the scale of the variations in the continuum. Second we fit a Gaussian
of fixed width equal to the variations in the continuum. We then step the Gaussian 
over the emission line free region pixel by pixel. Due to the fluctuations in 
the noise the Gaussian fit will sometimes have a positive normalisation 
and sometimes a negative one. The distribution of the Gaussian area values
in this region provide a second mechanism for estimating the one sigma
uncertainty. The errors determined in both fashions were consistent with each other.
Fig. \ref{all} shows
the upper limits overplotted on the continuum subtracted spectrum for the
[Fe {\small XIV}]$\lambda5303$, [Ca {\small XV}]$\lambda5445$, 
[Ca {\small XV}]$\lambda5694$, and [Ni {\small XV}]$\lambda6702$ lines.

\begin{table*}
 \centering
\begin{tabular}[h]{|c|c|c|c|c|c|c|}
\hline
\hline
Line & \multicolumn{3}{c}{Luminosity (\ergps)} & \multicolumn{3}{c}{Model (\ergps)} \\
     & Box 1.$^{a}$ ($10^{37}$) & Box 2.$^{a}$ ($10^{37}$) & 20 arcsec$^{2}$.$^{b}$ ($10^{37}$)
 & $Z=Z_{\odot}$ ($10^{37}$) & $Z=0.5Z_{\odot}$ ($10^{37}$) & $Z=Z_{\odot}$ ($10^{37}$) \\
\hline
$[\mathrm{Fe}$~{\scriptsize XIV}$]\lambda5303$ &$<$4.5&$<$4.0&$<$11.5&3.1&3.1&10.3\\
$[\mathrm{Ca}$~{\scriptsize XV}$]\lambda5445$ &$<$1.0&$<$0.9&$<$9.1&1.4&1.2&4.7\\
$[\mathrm{Ca}$~{\scriptsize XV}$]\lambda5694$ &$<$3.8&$<$1.4&$<$3.2&2.1&1.8&7.0\\
$[\mathrm{Fe}$~{\scriptsize X}$]\lambda6374$ &3.5$\pm$0.56&$<$7.0&25.8$\pm$8.3&4.1&4.2&13.7\\
$[\mathrm{Ni}$~{\scriptsize XV}$]\lambda6702$ &$<$1.1&$<$0.2&$<$1.7&0.2&0.2&0.6\\
\hline
\end{tabular}
\caption{The 90 per cent (2.7$\sigma$) upper limits on the luminosity
and detections of [Fe {\small X}]$\lambda$6374 emission
with one sigma errors in the regions specified. The luminosity quoted as detections 
used the continuum subtraction technique which gave the lowest integrated flux, in all
cases this was when using the SSP model fits. Where the results are upper limits and 
not clear detections the continuum subtraction method yielding the largest 
integrated flux was taken.
The model
approximate luminosities are from \protect \cite{sarazin1991} and scale 
simply with the cooling rate. Here they are given for a cooling rate of 3
(Columns 5 and 6) and 10
$\mathrm{M}_{\odot}~\mathrm{yr}^{-1}$ (Column 7). These limits are determined by spectral 
fitting of the X-ray spectrum \protect \citep{sanders2008b}. 
The sizes of the regions in square arc seconds (\kpc) are $^{a}$ 3.4 
arcsec$^{2}$ (0.7 \kpc$^{2}$) and $^{b}$ 20.1 arcsec$^{2}$ (4.2 \kpc$^{2}$). These
limits have been corrected for galactic extinction but not for the intrinsic 
reddening of NGC 4696. The upper limits on [Ca {\small XV}]$\lambda$5694
imply this is deficient by a factor of 2 in the inner 20 arcseonds compared with
the luminosity predicted from the X-ray cooling rate within this region.}
\label{results}
\end{table*}

\begin{table*}
 \centering
\begin{tabular}[h]{|c|c|c|c|}
\hline
\hline
Line & & Flux (\ergpcmsqps) & \\
     & Box 1.$^{a}$ & Box 2.$^{a}$ & All pixels.$^{b}$ \\
\hline
$[\mathrm{Fe~XIV}]\lambda5303$ &$<1.9\times10^{-16}$&$<1.7\times10^{-16}$&$<4.9\times10^{-16}$ \\
$[\mathrm{Ca~XV}]\lambda5445$ &$<4.4\times10^{-17}$&$<3.7\times10^{-17}$&$<3.9\times10^{-16}$ \\
$[\mathrm{Ca~XV}]\lambda5694$ &$<1.6\times10^{-16}$&$<5.9\times10^{-17}$&$<1.4\times10^{-16}$ \\
$[\mathrm{Fe~X}]\lambda6374$ &$1.5\times10^{-16}\pm2.4\times10^{-17}$&$<3.0\times10^{-16}$&$1.1\times10^{-15}\pm3.5\times10^{-16}$ \\
$[\mathrm{Ni~XV}]\lambda6702$ &$<4.6\times10^{-17}$&$<6.6\times10^{-18}$&$<6.0\times10^{-17}$ \\
\hline
\end{tabular}
\caption{Same as Column 1-4 of table \ref{results} in units of flux.}
\label{results2}
\end{table*}

\subsection{Spatial distribution of the hot gas}

The detection of [Fe {\small X}]$\lambda$6374 emission in box 1 but not
in box 2 implies there may be some `clumpiness' to the spatial distribution of
the intermediate temperature gas. A small change in temperature
equates to a vast change in emissivity (see Ferland et al. in prep), so this clumpy appearance may be indicative
of small variations in temperature of the gas across the field of view.

Using IFS observations we can trace the morphology of the regions where there appears
to be an excess of flux towards the red wing of the [O {\small I}]$\lambda$6363
line. Fig. \ref{flux} shows the spatial distribution of this excess flux across the
field of view, overlaid with the contours of radio emission (see 
\citealt{taylor2007}).
We binned the spectra in 5 by 5 regions, this size binning was used as a compromise
between spatial resolution, signal to noise and goodness of fit to the stellar continuum.
We then fit each region with SSP models as decribed
in section \ref{sec:subtraction}. The [O {\small I}]$\lambda$6363 object emission line was
then fit and removed and the remaining `counts' between $6363-6380~\mathrm{\text{\AA}}$
were summed (no emission line shape was assumed here) to create the map of relative 
flux, seen in the bottom panel of Fig. \ref{flux}. As an illustration we show, in 
Fig. \ref{bins}, spectra from bin A and B, the regions of highest excess flux, and 
spectra from bin C and D, where little excess flux is seen.

\subsection{Velocity of the hot gas}

Assuming a Gaussian profile for the [Fe {\small X}]$\lambda$6374 emission line detected in
box 1 we can fit the velocity width of the detected broad excess flux. We do this 
for each of the methods of continuum subtraction, as described above. The FWHM velocity width of
the feature determined from the different continuum subtraction methods is 300 to 650~\kmps, from box 1 (the
velocity width of the [O {\small I}]$\lambda$6363 is $\sim200$~\kmps). It 
should be noted that the predicted integrated line profile of the coronal line emission
is model dependent \citep{sarazin1991}, however our data are not deep enough to 
allow for an investigation of the line profile. The thermal line
width of the coronal line emission is typically only $20-30$~\kmps, so the broad width
of this feature is likely due to turbulent motions of the hot gas.
\cite{sarazin1991} show that for a gas of 2 million K with maximal turbulent broadening
(velocities limited by the sound speed in the hot ambient medium) predicted velocity
widths at FWHM are approximately 1700~\kmps. The motions in the gas we observe are 
therefore highly subsonic.

\section{Discussion}

We have conducted a deep search for 5 species of optical coronal
line emission, specifically [Fe {\small XIV}]$\lambda5303$, 
[Ca {\small XV}]$\lambda5445$, [Ca {\small XV}]$\lambda5694$, 
[Fe {\small X}]$\lambda6374$ and [Ni {\small XV}]$\lambda6702$.
These lines probe gas at temperatures of 2, 5, 5, 1.3 and 2.5 million K
respectively.

We report four upper limits 
and a 6.3 sigma detection of
[Fe {\small X}]$\lambda$6374 emission. This detected feature is broader than the
H$\alpha$ nebulosity at 10$^{4}$ K, having a FWHM velocity width of about 300$-$650~\kmps
and is blueshifted with respect to this cooler emission.
[Fe {\small X}]$\lambda$6374 is emitted from gas at one million K, the lowest
temperature of the coronal lines in our wavelength range.

Assuming that coronal line emission is a product of the low level of
residual cooling out of the hot gas observed in the central galaxies
of galaxy clusters we can determine the rate of cooling from the flux of
these lines. Upper limits for the detection of [Fe {\small XIV}]$\lambda$5303 and 
[Ni {\small XV}]$\lambda6702$ are consistent with the X-ray cooling rate, 
in the absence of heating, of $\sim$10~\Msunpyr\ \citep{sanders2008b} above
0.8~\keV, over
a region 20 arcseconds in diameter. However in the case of 
[Fe {\small XIV}]$\lambda$5303 we stress the difficulties involved in 
the subtraction of the stellar continuum which introduces a larger
error into the detection limit (see section \ref{sec:subtraction} and Fig.
\ref{blue}). 

The rate of cooling implied by the detection of [Fe {\small X}]$\lambda$6374
emission is a rate of $\sim$~20\Msunpyr\ in the central 20 arcsec$^2$, 
twice that inferred from X-ray observations and from the upper limits of
the other coronal lines (see table \ref{results}). If this emission is solely due to the 
cooling of the ICM in this cluster we would expect similar or 
greater detections from the [Fe {\small XIV}]$\lambda$5303 and [Ca {\small XV}]
lines. An alternative for the iron emission is that it originates in gas which
was cooler and has been heated.
There are lots of possible sources of heating including shocks, photoionisation from
central active galactic nuclei (AGN), thermal conduction, turbulent mixing or
heating associated with the radio bubbles. We caution that recent changes in the atomic parameters
indicate that the emissivity of [Fe {\small X}]$\lambda$6374 is larger than 
previously thought. This would imply a slightly lower cooling rate than calculated here
and will be investigated in further work by Ferland et al. (in prep).

C {\small IV} 1549$\mathrm{\text{\AA}}$
emission has been discovered by \cite{sparks2009} in the outer regions of
M87, co-located with its H$\alpha$ filaments. This probes the slightly
lower temperatures of $\sim$10$^{5}$~K. O {\small VI} 1032, 1035 $\mathrm{\text{\AA}}$ emission
($\sim$10$^{5.5}$~K) has also been seen in a small sample of cool core clusters
\citep{oegerle2001, bregman2006}. \cite{bregman2006} note that in Abell 1795
the cooling rate implied by the 10$^{5.5}$~K is larger than that implied by the 
X-ray emission, as is the case here. They suggests that non-steady cooling 
of material may be the cause of this conflict.

Most AGN show emission from forbidden, high-ionisation lines, the so
called extended coronal line regions (CLRs) \citep{mullaney2009, mazzalay2010}.
This emission is believed to be from a region outside of the broad-line 
region and inside the narrow-line region and lines are often observed to be
blueshifted by 500 or more \kmps. \cite{mazzalay2010} find that photoionisation
from the central AGN is the likely major ionisation mechanism for these
regions. Their results also confirm the observation that higher-ionisation
lines are emitted from a more compact area while the lower-ionisation lines 
can be found further into the narrow-line region. 

\begin{figure}
 \begin{center}
  \includegraphics[width=0.5\textwidth]{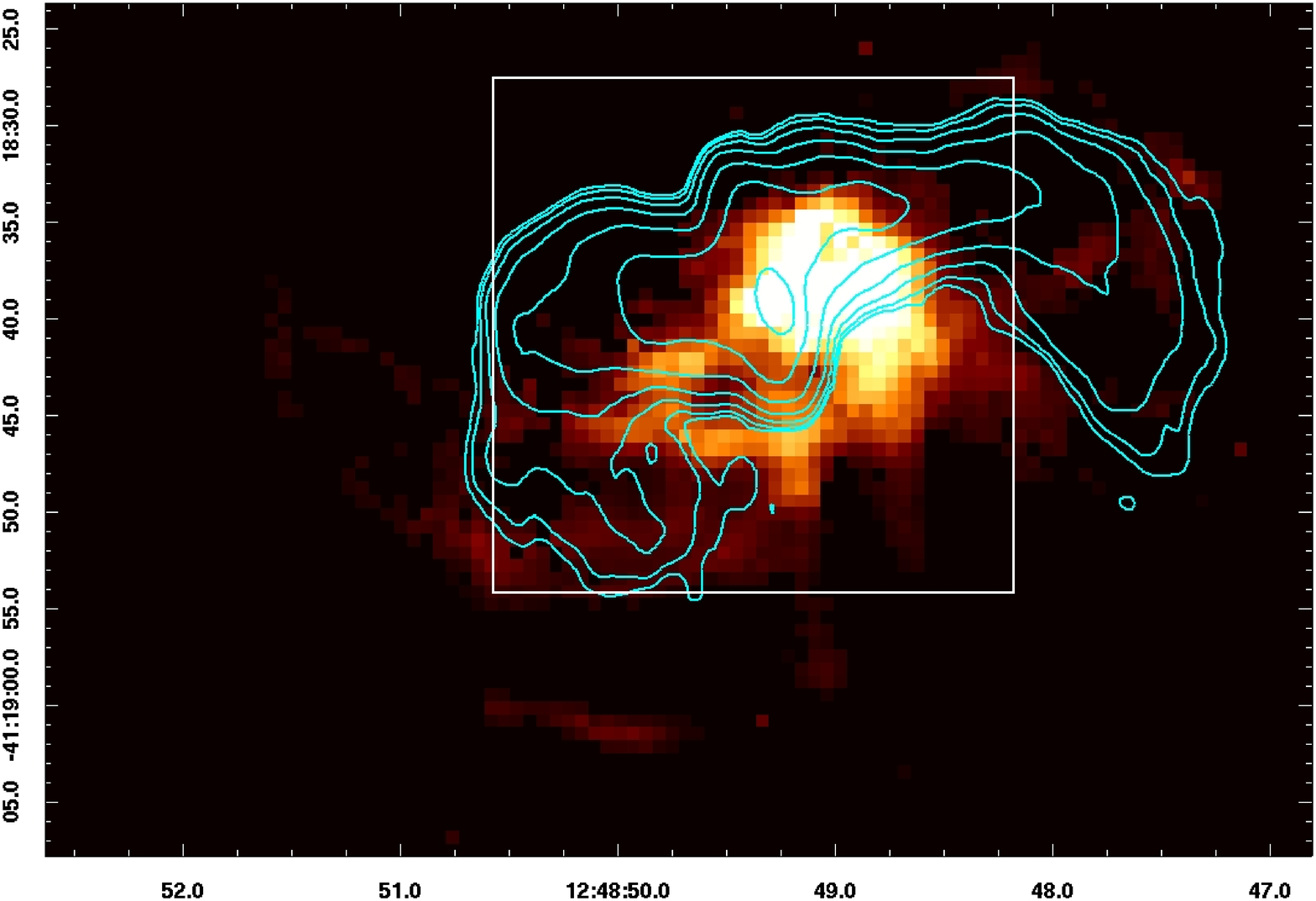}
  \includegraphics[width=0.5\textwidth]{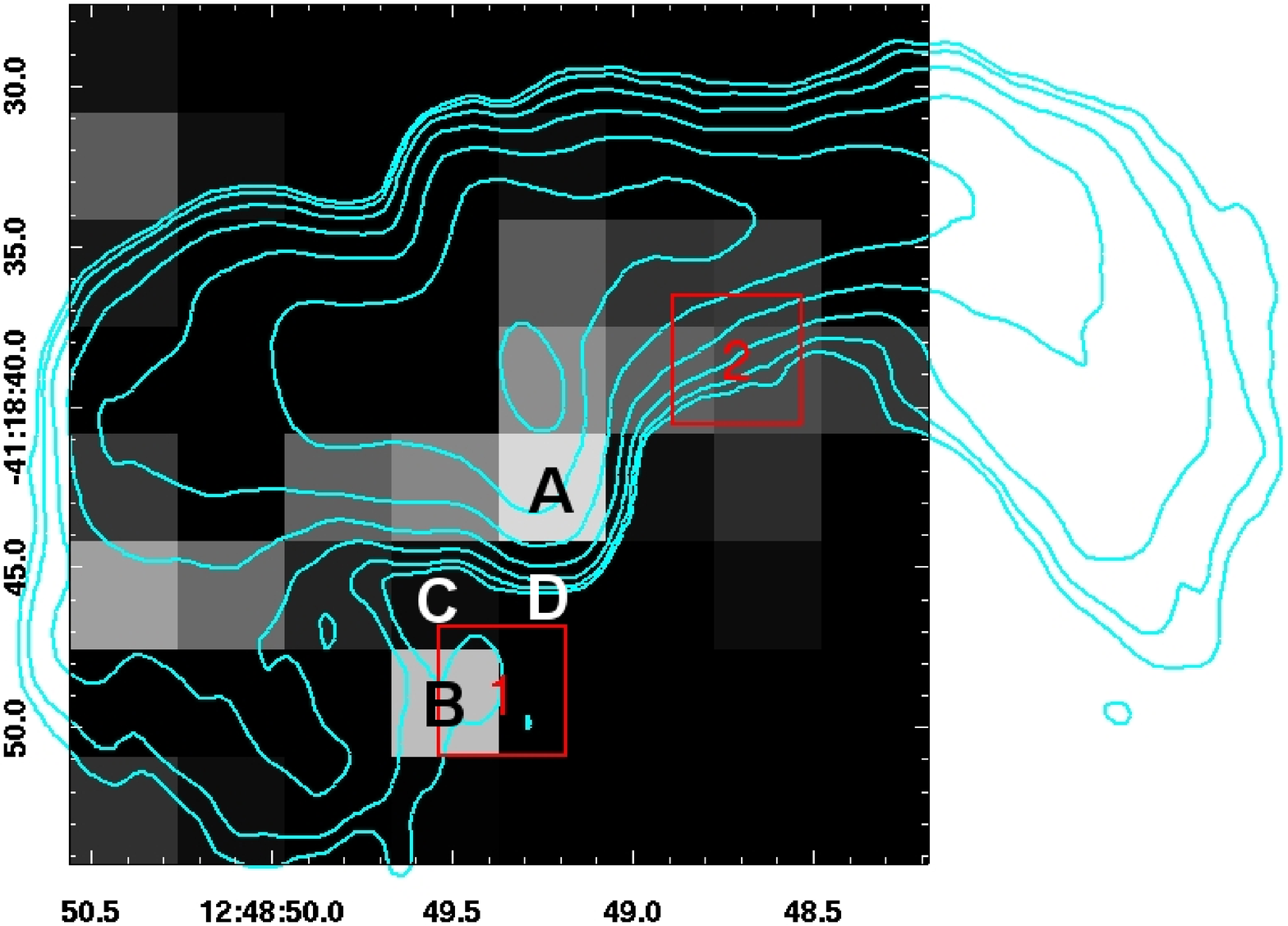}
\caption{The spatial distribution of the 1 million K gas. Above: The [N {\small II}]$\lambda$6583
emission, overlaid with VLA 5 \GHz\ radio contours 
(\citealt{taylor2007}. The white box indicates the region
where we have our deepest observations. Below: This region binned on a 5 by 5 fibre
basis showing the residual flux at the position of the [Fe {\small X}]$\lambda$6374 emission across
the field of view. The brightest bins are regions of highest flux. The spectra from
bins A-D are shown in Fig. \ref{bins}. For comparison box 1 and 2 are marked in red. \label{flux}}
 \end{center}
\end{figure}

\begin{figure}
 \begin{center}
  \includegraphics[width=0.45\textwidth]{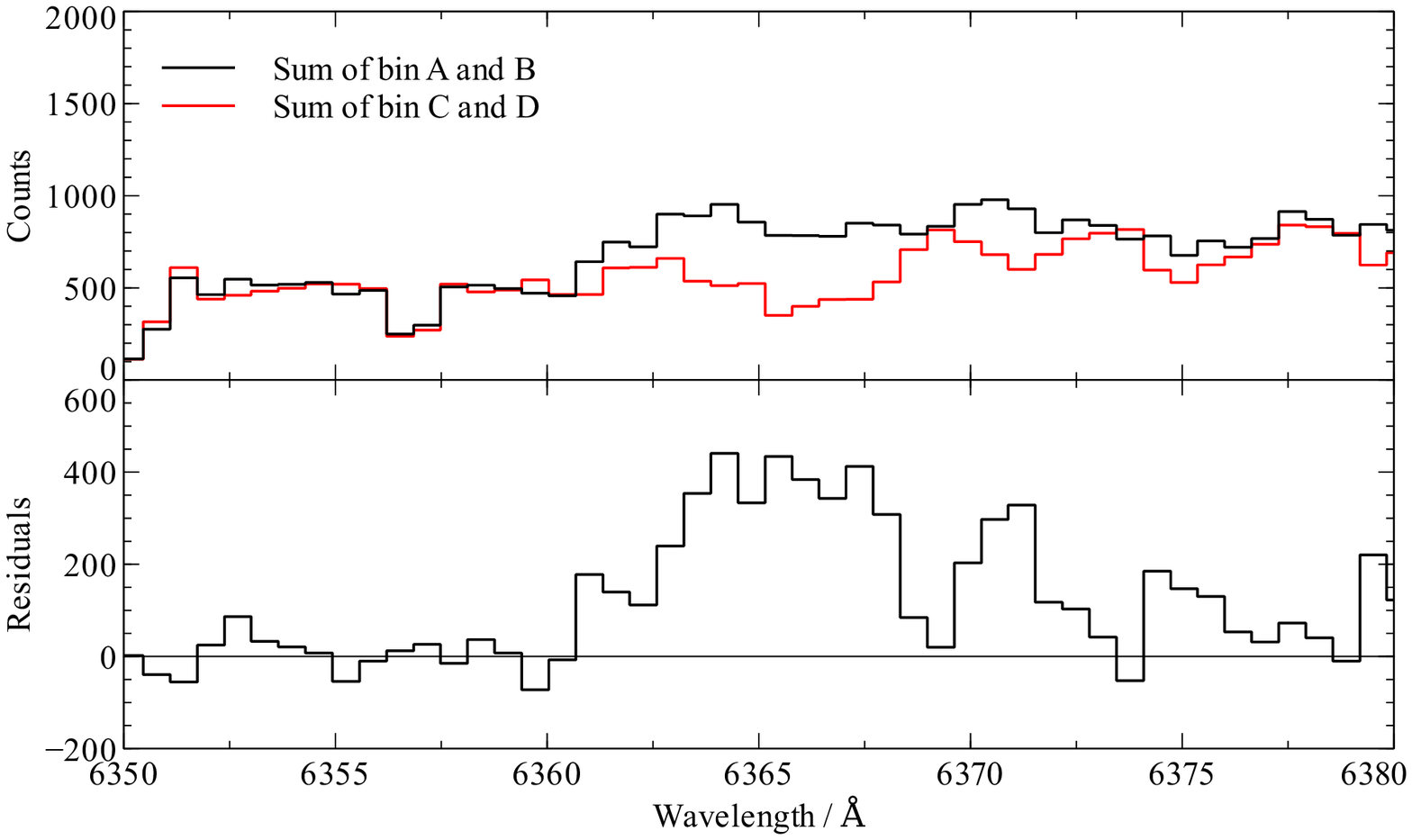}
\caption{A comparison of the spectra from four of the regions marked on Fig. \ref{flux}. The
spectra have been continuum subtracted and the [O {\small I}]$\lambda$6363 object 
emission line and the two sky lines to the left of the Oxygen line have been fitted and removed. 
The bottom panel shows the residuals between the two spectra. \label{bins}}
 \end{center}
\end{figure}

For a typical Seyfert
galaxy with an ionising luminosity $L_{\mathrm{ion}}=10^{43.5}$~\ergps, 
\cite{ferguson1997} find that the region containing [Fe {\small X}]$\lambda$6374 
emission should be restricted to the inner 20~\pc.
The size of the CLR scales with the ionising luminosity of the central source as 
$L_{\mathrm{ion}}^{1/2}$ \citep{ferguson1997}. Box 1, where we find significant
[Fe {\small X}]$\lambda$6374 emission, is located $\sim$15'' from the centre
of the radio source, a distance of $\sim$3~\kpc. Scaling up from the values of 
\citep{ferguson1997}, a region of the size $\sim$3~\kpc\ implies a huge ionising
luminosity of 10$^{47}$~\ergps. \cite{taylor2006} have found the nucleus of
NGC 4696 to be very faint. Their upper limit on the luminosity from X-ray 
observations between $0.1-10$~\keV\ is $\sim$10$^{40}$~\ergps. The spatial
extent over which we find the coronal line emission leads us to rule out the
possibility of photoionisation by the central AGN. It is possible, looking at
the residual flux over the spectral region where we would expect [Fe {\small X}]$\lambda$6374 emission,
that the spatial distribution of the 1 million K gas 
(Fig. \ref{flux}) traces that of the 10$^{4}$~K filaments 
and large dust lane however this is a tentative result and would need confirmation
with deeper observations.

\begin{figure}
 \begin{center}
  \includegraphics[width=0.45\textwidth]{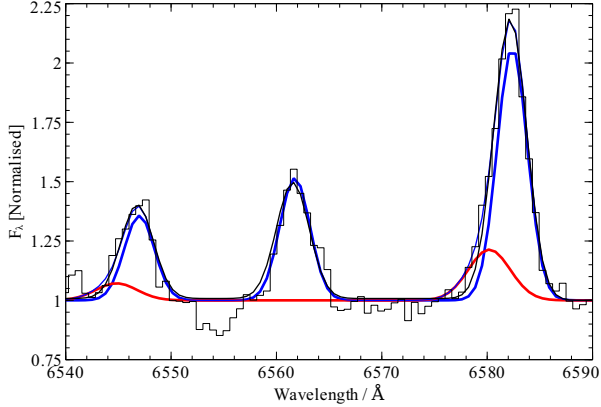}
\caption{A two component fit to the H$\alpha$ and [N {\small II}] emission from box 1.
There is a evidence for a slightly blueshifted component to these lines with a blueshifted
velocity of $\sim$140\kmps. This is smaller than the velocity shift seen in the 
[Fe {\small X}]$\lambda$6374 emission line of $\sim$230\kmps. \label{ha_nII}}
 \end{center}
\end{figure}

Another explanation for the excess flux such as emission from
a merging clump of gas or another filament is unlikely, as any extra component 
of [O {\small I}]$\lambda$6363 emission should also be present in the 
[O {\small I}]$\lambda$6300 emission and no red wing is observed on this line.
One explanation for the blueshift could be that the radio source is interacting 
with and driving the plasma away possibly entraining with it cooler gas.  
Fig. \ref{ha_nII} shows a two component fit to the [N {\small II}] and H$\alpha$
emission in the box 1 spectrum. There is evidence for a blueshifted component
to these lines and the best fit gaussians imply a blueshifted velocity of
$\sim$140\kmps. This is less than the observed blueshift of the 
[Fe {\small X}]$\lambda$6374 line ($\sim$230\kmps) so is inconclusive.

\cite{sanders2008b} have used deep \xmm\ Reflection Grating Spectrometer
(RGS) observations to show that the centre of the Centaurus cluster
contains X-ray emitting gas down to at least 4 million K through
detections of the Fe {\small XVII} line and limits on the O {\small VII} emission.
The metal abundances are very high in the inner 30~\kpc\ of this 
cluster \citep{sanders2006}. The iron abundance of the
hot gas is $\sim$2 times solar and the calcium 
abundance $\sim$3 times solar. The \chandra\ spectra show evidence of an
off-centre abundance peak, at $\sim$20~\kpc. The best traced elements
of Fe and Si then show a decline in abundance in the nucleus.

At these high abundance values we expect to detect, in our optical spectra, 
lines of [Ca {\small XV}], probing gas of 5 million K, detected already 
by its X-ray Fe {\small XVII} emission. Contrary to this expectation
we do not detect any significant calcium emission. Upper limits 
in most cases suggest the rate of radiative cooling is less than that inferred from 
the X-ray spectra of hotter gas and provide a more stringent constraint than the other
coronal emission lines. 

We explore four possible explanations for this result; inaccurate continuum 
subtraction; a lower cooling rate; a low calcium abundance and 
calcium deficiency in the intermediate temperature gas. We will deal
with each of these points in turn.

The continuum subtraction has been attempted in a variety of ways (see section 
\ref{sec:subtraction}) and in all cases produces similar results. Where
these results were upper limits and not clear detections the largest value was
taken and is given in table \ref{results}. Where the stellar spectra exhibit 
many features the uncertainties are large and the detection of broad line emission is
very difficult. However, in the absence of a template spectrum in which we 
are sure there is no emission from intermediate temperature gas and observations 
with similar abundance ratios across the field of view this is the best we 
can do.  

The cooling rates are derived for cooling without heating and so may be lower 
than stated, they may also vary across the temperature range probed. Investigations 
of the quantity of molecular gas and dust and of recent star formation in these
objects may help constrain the quantity of cooling, to very low temperatures, that we
expect. A much lower cooling rate may provide an explanation for the lack of [Fe {\small XIV}]$\lambda$5303
and [Ni {\small XV}]$\lambda$6702 emission we see however the calcium lines should 
originate from a temperature range that has been directly observed from 
measurements of the Fe {\small XVII} line ratios, so the lack of this emission
is not easily understood.

The Centaurus cluster has very high metallicities which peak at
a radius of 20~\kpc, then decline towards the nucleus \citep{sanders2006}. This enhancement and 
decline in the metallicities are reproduced using both \xmm, \chandra\ CCD 
observations and \xmm\ RGS observations. An off-centre abundance peak with a depression
in abundance towards the centre of clusters and groups is often observed in
objects with cool cores (see for example \citealt{degrandi2001, rasmussen2007} and 
\citealt{degrandi2009}). 

There are a number of biases and uncertainties in abundance measurements from 
spectral modelling. Inaccuracies in modelling the temperature structure of galaxy
clusters can introduce an `Fe bias' \citep{buote2000}. Here the measured Fe
abundance is lower than the actual value due to fitting a multiphase gas with
single or only a couple of temperature components. The opposite effect,
an `inverse Fe bias', is also seen in cool core clusters when a single temperature 
model is fitted to a multiphase plasma with temperatures between $2-4$~\keV\ 
\citep{rasia2008}. In the case of Centaurus the high spectral resolution of
\xmm\ RGS allows the temperature to be constrained with line strength ratios giving
a more robust check on the temperature components of the models \citep{sanders2008b}.

Sedimentation in the centre of galaxy clusters \citep{fabian1977b}
can cause the metal abundances to rise, and could be 
reversed by the effects of thermal diffusion \citep{shtykovskiy2010}.
Neglecting the effects of resonance scattering also underestimates the abundances
of metals in clusters. This effect is at most 10 per cent,
so cannot fully explain the central abundance dips observed in galaxy clusters \citep{sanders2006b}. The drop in 
abundances in the Centaurus cluster may be explained with a complex model 
involving three temperature components and additional absorption however 
the errors on the inner-most radial bins become sufficiently large that it is 
impossible to tell if the drop is real or not \citep{fabian2005b}.

A major contributor to a central abundance drop must be depletion on dust.
The central few \kpc\ of the hot gas will be dominated by stellar mass loss, in which most
metals are bound in dust grains. They will slowly be introduced into the hot phase
by sputtering, at a rate dependent on grain size \citep{draine1979, barlow1978}. Much iron 
however could be injected into the hot phase through SNIa.

Calcium
is one of the most depleted of all refractory elements in the presence of dust
\citep{field1974, spitzer1975, cowie1986}, with depletions of 10$^{-4}$
relative to solar abundance typical in dense clouds. NGC 4696 
hosts a huge dust lane which almost completely encircles the core and 
spirals out to the north east, tracing the morphology of the H$\alpha$ filaments. 
The X-ray absorption column density is also highest in the same 
region \citep{crawford2005}. It has moreover a large quantity of infrared-emitting dust in its nucleus  
\citep{goudfrooij1995, kaneda2007}. \cite{kaneda2007} show from
Spitzer Multiband Imaging Photometer (MIPS) 24~$\mu$m 
surface brightness profiles that the dust emission increases steeply in the inner 
30~arcseconds (6.3~\kpc), a region slightly larger than our field of view
and where we observe the metallicity drop.

The cumulative gas mass obtained from 
the X-ray emitting hydrogen density profile of \cite{graham2006} in the inner 30~arcseconds or 
6.3~\kpc\ (10~arcseconds, 2.1~\kpc) is $9.4\times10^{8}$~\Msun\ 
($9.2\times10^{7}$~\Msun). Assuming a dust to gas ratio similar to our 
Galaxy \citep{crawford2005, edge2010} we get a total dust mass of $9.4\times10^{6}$~\Msun\
($9.2\times10^{5}$~\Msun). The dust mass estimated by \cite{goudfrooij1995}
using the Infrared Astronomical Satellite (IRAS) is $4.7\times10^{6}$~\Msun\
and the Spitzer MIPS result from \cite{kaneda2007} is $4.2\times10^{5}$~\Msun. These
values are considered to be lower limits to the dust mass as the instruments
are insensitive to very cold dust and imply that the inner interstellar medium
in NGC 4696 is highly deficient in refractory elements.

A number of cool core clusters have been fround to be deficient in calcium in the warm (10$^{4}$~K) emission
line nebulae, where [Ca {\small II}] and [Ca {\small V}] emission lines would be expected
\citep{ferland1993, donahue1993}. Our 
spectra of NGC 4696 are consistent with this since they 
show no evidence for [Ca {\small II}] or [Ca {\small V}] emission in the inner regions 
of NGC 4696.

These results indicate that NGC 4696 is deficient in gas-phase calcium at temperatures
below $\sim$5 million K. The lack of this warm and intermediate temperature gas 
phase in the inner 20 arcseconds implies the calcium in dust has never been part of 
the hot ICM and was probably introduced in dusty stellar mass-loss.

The high calcium abundance in the hot gas, discovered by \cite{sanders2006}, has its
greatest contribution from the hot 4~\keV\ emission (the contribution from the best 
fit to the lower temperature component, 0.5~\keV, is negligible). This is consistent
with the dust being sputtered by the hotter gas at large radii ($\geq$few~\kpc), where
there are also no shielding `cold' filaments.

\section{Conclusion}

We report the detection of [Fe {\small X}]$\lambda$6374 coronal line emission in
NGC 4696. This emission probes temperatures of 1 million K and is the first detection 
of intermediate temperature gas in this object. We fail to detect emission
from coronal lines of higher temperature gas including those of [Ca {\small XV}]
which probe gas of 5 million K and which we had expected to detect due to the 
high abundance of calcium in the hot X-ray emitting gas.

We conclude that calcium is likely to be depleted in the dusty central regions
of NGC 4696. This is consistent with our apparent lack of [Ca {\small V}] and [Ca {\small II}]
ions which probe lower temperature gas and with the negligible contribution to the 
abundance in the hot gas of the lower temperature 0.5~\keV\ calcium lines. The dust in the central 
region of the galaxy is likely due to stellar mass-loss and has survived as dust grains due to the 
shelter of surrounding cooler gas. 
The abundance of calcium is higher in the outer 4~\keV\ ICM since dust is sputtered there 
where the dust is less protected by the cooler 
surroundings and the cold filaments. Deeper, high resolution X-ray observations 
which better constrain the central abundances of Fe, Si, Ca and Ne would help to
distinguish between the processes that contribute to the metal abundance.

There is now strong evidence to show that the central galaxies in many cool core 
clusters are playing host to large quantities of dust and cool gas (see for example 
\citealt{mcnamara1992, edge2010}). \cite{donahue1993} have also shown that  
the [Ca {\small II}] doublet is much weaker than expected in a sample of BCGs.
They conclude similarly that calcium is most likely depleted into dust grains in their
sample.   

The cooling rate inferred by the [Fe {\small X}]$\lambda$6374 emission is large
(20~\Msunpyr in a spatial region of 20 arcseconds$^{2}$) compared with that determined from other lines in the optical and
X-ray spectrum. This and the apparent lack of [Fe {\small XIV}]$\lambda$5303 
and [Ni {\small XV}]$\lambda$6702 emission rule out a steady cooling flow from the
10$^7$~K gas in this object. Some gas may however be cooling non-radiatively for
example by mixing with the colder gas. The strength of the [Fe {\small X}] emission
suggests that the million K gas is being heated rather than condensing out of the hot ISM. 

\section{Acknowledgements}
REAC acknowledges STFC for financial support. ACF thanks the Royal Society. GJF gratefully acknowledges 
support by NSF (0607028 and 0908877) and NASA (07-ATFP07-0124). REAC would also like to thank 
Rob Sharp for allowing use of his IFU IDL routines and Ryan Cooke, Paul Hewett and Ben Johnson for
help and valuable discussions.

This research has made use of the NASA/IPAC Extragalactic Database (NED) which is operated by the Jet 
Propulsion Laboratory, California Institute of Technology, under contract with the National Aeronautics 
and Space Administration.

The \Starlight\ project is supported by the Brazilian agencies CNPq, CAPES and
FAPESP and by the France-Brazil CAPES/Cofecub program.

The data published in this paper have been reduced using VIPGI, designed by the VIMOS Consortium 
and developed by INAF Milano.

The figures in this paper were produced using {\sc Veusz}\footnote[2]{http://home.gna.org/veusz/}.

\bibliographystyle{mnras}
\bibliography{/home/bcanning/Documents/latex_common/mnras_template}

\end{document}

%% file: defn.tex
\newcommand{\Mpc}{\rm\thinspace Mpc}
\newcommand{\kpc}{\rm\thinspace kpc}
\newcommand{\pc}{\rm\thinspace pc}
\newcommand{\km}{\rm\thinspace km}

\newcommand{\cm}{\rm\thinspace cm}

\newcommand{\yr}{\rm\thinspace yr}

\newcommand{\Myr}{\rm\thinspace Myr}
\newcommand{\s}{\rm\thinspace s}

\newcommand{\GHz}{\rm\thinspace GHz}

\newcommand{\Msun}{\hbox{$\rm\thinspace M_{\odot}$}}

\newcommand{\Msunpyr}{\hbox{$\Msun\yr^{-1}$}}

\newcommand{\keV}{\rm\thinspace keV}

\newcommand{\erg}{\rm\thinspace erg}

\newcommand{\ergpcmsqps}{\hbox{$\erg\cm^{-2}\s^{-1}$}}

\newcommand{\ergps}{\hbox{$\erg\s^{-1}\,$}}

\newcommand{\kmps}{\hbox{$\km\s^{-1}\,$}}

\newcommand{\kmpspMpc}{\hbox{$\kmps\Mpc^{-1}\,$}}

\newcommand{\asec}{\rm\thinspace arcsec}

\def\Starlight{\hbox{\sc STARLIGHT}}

\def\chandra{{\it Chandra}}
\def\xmm{{\it XMM-Newton}}

